\documentstyle[prl,aps]{revtex}
\begin{document}
\draft
\title{Completely Positive Quantum Dissipation}
\author{Bassano~Vacchini}
\address{Dipartimento di Fisica
dell'Universit\`a di Milano and INFN, Sezione di Milano,
Via Celoria 16, I-20133 Milan,
Italy
\\
Fachbereich Physik, Philipps-Universit\"at, Renthof 7,
D-35032 Marburg, Germany}
\date{\today}
\maketitle
\begin{abstract}
A completely positive master equation describing quantum
dissipation for a Brownian particle is derived starting from
microphysical collisions, exploiting a recently introduced
approach to subdynamics of a macrosystem.
The obtained equation can be cast into Lindblad form with a
single generator for each Cartesian direction.
Temperature
dependent friction and diffusion coefficients for both
position and momentum are expressed in terms of the
collision cross-section.
\end{abstract}
\narrowtext
The issue of quantum dissipation, and in particular of
quantum Brownian motion, is a long standing one, which
attracts physicists since decades (for general references
see~\cite{Dekker-Grabert,Alicki}) and still seems to
be unsolved. Its relevance however is growing, especially in
connection with decoherence and the relationship between
classical and quantum  description~\cite{Kiefer}, a field
which seems to be now within reach of experimental
tests~\cite{Haroche}.
The classical understanding of the phenomenon is quite well
established, relying on Langevin or Fokker-Planck equations obtained
considering a particle typically interacting with a bath of
independent oscillators, so that most of the research work
has been influenced by these results, amounting to a
research for a quantum analog or quantum generalization of
these equations. The difficulty lies in the failure of a
Hamiltonian description for such systems, so that a clear
quantization prescription is missing and one needs a
thoroughly quantum mechanical approach. The most promising
results come from the reduced description of a particle
interacting with some kind of reservoir, thus impinging on
techniques and results of open quantum system theory.
In this respect the property of  complete positivity (CP) has
emerged as a very useful and stringent requirement in the
study of subdynamics inside quantum
mechanics~\cite{Lindblad,Alicki}. The property
of CP amounts to ask that the time
evolution semigroup ${\cal U}(t)$ for the irreversible
dynamics has the structure
$
{\cal U}(t)\varrho = \sum_\alpha K_\alpha (t) \varrho
K^\dagger_\alpha (t)
$, ($\sum_\alpha K^\dagger_\alpha (t)K_\alpha (t) =1$),
so that in particular  positivity is preserved, and it
origins from the formal requirement that interactionless
coupling to an n-level system does not affect positivity.
Indeed CP appears somehow more natural if
considered in the context of operations and quantum
measurement in which it originally appeared in
physics~\cite{Kraus-Hellwig}.
According to a famous paper by Lindblad~\cite{Lindblad}
under suitable mathematical
conditions the property of CP allows for
the determination of the general structure of the generators of
irreversible time evolutions, even though a thorough
understanding of the physical limits of validity of this
property is still on its way~\cite{Piza,Pechukas}, so
that satisfaction of CP by itself does not
ensure physically meaningful results, and the connection to
realistic microphysical models is strongly desirable. In
fact recent work has stressed the connection between
CP of the time evolution and weak-coupling,
so that for example an uncorrelated statistical operator can
consistently be considered as initial condition, as well as
coarse-graining in time~\cite{Royer-Suarez}. Most of the
work carried out at a fundamental level regarding
dissipative  systems follows the Feynman-Vernon influence
functional formalism (see~\cite{Dekker-Grabert,Hu} and
references therein),
first used by Caldeira and Leggett~\cite{Caldeira}, who have
derived a standard model of quantum Brownian motion, which
however has the drawback that it does not ensure the
positivity of the statistical
operator~\cite{Ambegaokar-Pechukas2}. This shortcoming has been
cured in various ways~\cite{Hu,Diosi1-Diosi2} and within the
independent oscillator model also a positive, though not
CP time evolution has recently been
obtained~\cite{Sipe}. On the side of  CP
time evolutions very little has been done at a fundamental
level, thus also leading to the conjecture that
CP dissipative evolutions could not be obtained from
first principles~\cite{Munro}, while most approaches are
either axiomatic or phenomenological~\cite{pheno}, so
that few insights about what kind of physical systems might
consistently be described by Lindblad-type master equations
can be gained.
\par
In this Letter we
give a derivation of
CP quantum Brownian evolution based on
some recent work on the description of coherent and
incoherent particle-matter interaction~\cite{art1}, which
has already been successfully applied to neutron
optics~\cite{art2}. This work relies on the introduction of
a time scale for the description of the subdynamics of
slowly varying degrees of freedom in the Heisenberg picture,
in this specific case the heavy particle with respect to the
background of thermal particles. The interaction is in terms
of two-particle collisions described by the full T-matrix
and the structure ensuring CP naturally
arises from the resolvent identity of scattering theory.
%%%%%%%%%%%%%%%%%%%%%%%%%%%%%%%%%%%%%%%%%%%%%%%%%%%%%%%%%%%%%
\paragraph*{Derivation and structure of the master
equation.}
%%%%%%%%%%%%%%%%%%%%%%%%%%%%%%%%%%%%%%%%%%%%%%%%%%%%%%%%%%%%%
Let
$
        {H}={H}_0 + {H}_{\text{m}} + {V}
$
be the Hamiltonian of the whole confined system
in second quantization,
$
        {H}_0 = \sum_h
        {E_h} {a^{\scriptscriptstyle \dagger}_{h}} {a_{{h}}}
$
describing the particle (either a fermion or a boson in the
state $u_h$), ${H}_{\text{m}}$ matter and $V$ their mutual
interaction. We intend to describe a single particle, so
that we consider for the total system the statistical
operator
$
        {\varrho}=  
        \sum_{{k} {h}}{}  
        {a^{\scriptscriptstyle \dagger}_{k}}  
        {{\varrho}^{\text{m}}} {a_{h}}
        {{\varrho}}_{kh}
$, where ${{\varrho}^{\text{m}}}$ describes matter,
${{\varrho}}_{kh}$ is a positive matrix with trace one
and we have
$
        {a_{{h}}}{{\varrho}^{\text{m}}}=0
$,
$
        {{\varrho}^{\text{m}}}  
        {a^{\scriptscriptstyle \dagger}_{h}}=0
$,
$
        \forall h  
$.
In order to consider the subdynamics of the microsystem
we exploit the following reduction formula, which connects
the expectations of operators in the total Fock space
${{{\cal H}_{\scriptscriptstyle F}}}$ with those of
operators in the one particle Hilbert space
${{\cal H}^{(1)}}$
        \[  
        {\hbox{\rm Tr}}_{{\cal H}_{\scriptscriptstyle F}}  
        \left(  
        {{A}{\varrho}}  
        \right)  
         = \sum_{h,k} {A}_{hk}  
        {{\varrho}}_{kh}=  
        {\hbox{\rm Tr}}_{{{\cal H}^{(1)}}}  
        \left(  
        {{\hat {\sf A}} {\hat {\varrho}}}  
        \right)
        \]  
where $A$ has the typical structure
${A}  
        = \sum_{h,k}  
        {a^{\scriptscriptstyle \dagger}_{h}}  
        {A}_{hk}  
{a_{k}}  
        = \sum_{h,k}  
        {a^{\scriptscriptstyle \dagger}_{h}}  
\langle  
h  
\vert  
{\hat {\sf A}}  
\vert  
k  
\rangle  
{a_{k}}$.
We intend to work on a time scale $\tau$ much
longer than microphysical collision time, being only
interested in the slow dynamics of the particle, and we
shall therefore approximate the time derivative with the
following coarse-grained one:
        \begin{eqnarray*}
        {  
        \Delta_{\tau} {\varrho}_{kh}(t)  
        \over  
        \tau  
        }  
        &=&
        {1\over \tau}  
        \left[  
        {\varrho}_{kh}(t+\tau) -  
        {\varrho}_{kh}(t)  
        \right]  
        \\
        &=&
        {1\over \tau}  
        \left[  
        {\hbox{\rm Tr}}_{{\cal H}_{\scriptscriptstyle F}}  
        \left(  
        {a^{\scriptscriptstyle \dagger}_{h}} {a_{k}}  
        e^{-{{  
        i  
        \over  
         \hbar  
        }}H\tau}  
        \varrho (t)  
        e^{{{  
        i  
        \over  
         \hbar  
        }}H\tau}  
        \right)  
        -  
        {{\varrho}}_{kh}(t)  
        \right]  
        .  
        \end{eqnarray*}
Exploiting the cyclic invariance of the trace we are led to consider
H-picture operators
$
{{a^{\scriptscriptstyle \dagger}_{h}}(\tau){a_{k}}}(\tau)
$,
to be evaluated on the given time scale using
a superoperator formalism, so that e.g.
${\cal H}={i \over \hbar} [{H},\cdot]$, using the
integral representation
        \[  
        a_{k}(\tau)
        =  
        {{e^{{\cal H}\tau}}{a_{k}}}
        =
        {\int_{-i\infty+\eta}^{+i\infty +  \eta}}{
        dz  
        \over  
            2\pi i  
        }       \,   e^{z \tau}  
        {  
        {{  
        \left(  
        {{ z - {\cal H}}}
        \right)  
        }^{-1}}  
        {a_{k}}}  
        .  
        \]  
Introducing the superoperator
$
        {\cal T}(z)  
        \equiv  
        {\cal V} + {\cal V}{{
        \left(  
        {{ z - {\cal H}}}
        \right)  
        }^{-1}}{\cal V},
$
which is the analog of the T-matrix, we have
        \[  
        {{  
        \left(  
        {{ z - {\cal H}}}
        \right)  
        }^{-1}}={{  
        \left(  
        {{ z - {\cal H}_0}}
        \right)  
        }^{-1}} +{{  
        \left( 
        {{ z - {\cal H}_0}}
        \right) 
        }^{-1}}  
        {\cal T}(z){{  
        \left(  
        {{ z - {\cal H}_0}}
        \right)  
        }^{-1}}  
        \]  
so that in the considered structure
$
{{a^{\scriptscriptstyle \dagger}_{h}}(\tau){a_{k}}}(\tau)
$, bilinear in the field operators, the emergence of a
typically incoherent term having the CP
structure
$K_\alpha \varrho K^\dagger_\alpha$ bilinear in the T-matrix
naturally appears, thus confirming recent phenomenological
approaches~\cite{Stegholm,Diosi}. Using the fact that
$[{H},\sum_h {a^{\scriptscriptstyle \dagger}_{h}} {a_{{h}}}  ]=0$
(the interaction potential being bilinear in the particle field operators)
the restriction of ${{\cal T}(z)}{a_{k}}$ to the case of 
a single particle may be generally written
$
        i\hbar  
        {{{\cal T}(z)}{a_{k}}}
        =\sum_h  
        T{}_{h}^{k}  
        \left(  
         i\hbar  z  
        \right)  
        {a_{{h}}} ,  
$
where $  
T{}_{h}^{k}  
\left(  
   z  
\right)  
$ is an operator in the Fock space of the macrosystem only.
This matrix, which according to the introduction of the time
scale should exhibit a slow energy dependence, plays a
central role, accounting for the peculiarities of the
interaction between the particle and the considered medium.
The master equation we finally obtain has the following
Lindblad form
        \begin{equation}
        \label{eq}
        {  
        d {\hat \varrho}  
        \over  
                      d\tau  
        }  
        =
        -
        {i \over \hbar}
        [{\hat {{\sf H}}}
        ,
        {\hat \varrho}  
        ]
        +
        {1\over\hbar}
        \sum^{}_{{\lambda,\xi}}
        \bigl[
        {\hat {\sf L}}_{\lambda\xi} {\hat {\varrho}}
        {{\hat {\sf L}}{}_{\lambda\xi}^{\scriptscriptstyle  
        \dagger}}  
        -
        {\scriptstyle {1\over 2}}  
        \{
        {  
        {{\hat {\sf L}}{}_{\lambda\xi}^{\scriptscriptstyle \dagger}}  
        {\hat {\sf L}}{}_{\lambda\xi}  
                , {\hat \varrho}  
                }
        \}
        \bigr]
        ,
        \end{equation}
where
$
{\hat {{\sf H}}}= {\hat {{\sf H}}}_0 +
 {\hat {\sf V}}
$ and
$
        {\hat {\sf V}}=
        \frac 12
        (
        {{\hat {\sf Q}}+
        {\hat {\sf Q}}{}^{\scriptscriptstyle \dagger}}  
        )
$  with
        \begin{eqnarray*}
        \langle  
        {k}
        \vert  
        {\hat {\sf Q}}  
        \vert  
        {h}  
        \rangle  
        &=&
        {\hbox{\rm Tr}}_{{{{\cal H}_{\scriptscriptstyle F}}}}
        \left[
        {
        {
        T{}_{h}^{k}
        ({E_k}+i{\varepsilon})
        }
        {{\varrho}^{\text{m}}}
        }
        \right]
        \\
        \langle  
        {k}
        \vert  
        {\hat {\sf L}}_{\lambda\xi}
        \vert  
        {h}  
        \rangle  
        &=&
        \sqrt{2\varepsilon  \pi_\xi}
        {
        \langle
        \lambda
        \vert
        {
        T{}_{h}^{k}
        ({E_k}+i{\varepsilon})
        }
        \vert
        \xi
        \rangle
        \over
        {{E_k}+{E_{{\lambda}}}-{E_h}-{E}_{\xi} -i\varepsilon}
        }
        ,
        \end{eqnarray*}
being
$
{{\varrho}^{\text{m}}}=\sum_\xi \pi_\xi
\vert\xi\rangle\langle\xi\vert
$
the statistical operator  describing matter at equilibrium
and
$\vert\lambda\rangle,
\vert\xi\rangle$ denoting eigenvectors of
${H}_{\text{m}}$ with eigenvalues $E_\lambda, E_\xi$, while $\vert k\rangle,
\vert h\rangle$ denote eigenvectors of
${H}_0$ with eigenvalues $E_k, E_h$. A
detailed derivation of this master equation is given
in~\cite{art1,art2}.
In considering many-particle systems important corrections due to
statistics of identical particles appear. This will not be the case
here since we consider a single particle distinguished from its
environment through the Brownian limit in which the ratio between the
masses is very far from unit.
A generalization of the formalism to cope with dilute many-particle 
systems, in which statistic effects have been accounted for, 
has been considered in Ref.~\cite{berlin-goslar}, 
aiming at understanding the transition between quantum and
classical .
%%%%%%%%%%%%%%%%%%%%%%%%%%%%%%%%%%%%%%%%%%%%%%%%%%%%%%%%%%%%%
\paragraph*{Completely positive quantum Brownian motion.}
%%%%%%%%%%%%%%%%%%%%%%%%%%%%%%%%%%%%%%%%%%%%%%%%%%%%%%%%%%%%%
Following~\cite{art2} we will make the general Ansatz
$
        T{}_{h}^{k} \left( z \right) 
        =
        {\int d^3 \! {\bbox{x}} \,}
        {\int d^3 \! {\bbox{y}} \,}  
        \psi^{\scriptscriptstyle\dagger}({\bbox{x}})  
        u_k^{*}({\bbox{y}})  
        t(z,{\bbox{x}}-{\bbox{y}})  
        u_h({\bbox{y}})  
        \psi({\bbox{x}})  
$,
where we have supposed translation invariance in the
interaction kernel, and $\psi^{\scriptscriptstyle\dagger}$,
$\psi$ denote field  operators for the macrosystem.
Introducing creation and destruction operators
$b^{\scriptscriptstyle\dagger}$, $b$ in the Fock space of the
macrosystem, we may write
$
        T{}_{h}^{k} \left( z \right)
        =  
        \sum_{\eta  \mu}
        b^{\scriptscriptstyle\dagger}_{\eta}
        T_{k\eta h \mu}(z)
        b_\mu  
$.
Being interested in local dissipation effects we may safely
suppose that at least far away from the boundaries the
system is homogeneous so as to use as quantum numbers
momentum eigenvalues, thus obtaining
$
        T_{k\eta h \mu}(z)
        =
        \delta_{p_\eta +p_k,p_h+p_\mu}
        \tilde{t} (z,
        \left |
        {\bbox{p}}_{\mu}-{\bbox{p}}_{\eta}
        \right |
        )
$, depending on the Fourier transform of the interaction
kernel, and therefore
        \begin{equation}
        \label{ansatz}
        T{}_{h}^{k}  (z)
        =
        \sum_{\eta\mu}
        \delta_{p_\eta +p_k,p_h+p_\mu}
        b^{\scriptscriptstyle\dagger}_{\eta}
        \tilde{t} (z,
        |
        {\bbox{p}}_{\mu}-{\bbox{p}}_{\eta}
        |
        )
        b_\mu                      .
        \end{equation}
We may now insert this expression in (\ref{eq})
to
evaluate the different contributions, starting from the
last, typically incoherent term. In doing this we consider
the medium as composed of free gas particles, so that the
energy eigenstates $\vert\lambda\rangle$,
$\vert\xi\rangle$ of ${{\varrho}^{\rm m}}$ may
be obtained by the repeated action of
$b^{\scriptscriptstyle\dagger}_l$ on the vacuum, and we can
write
$
        \vert\lambda\rangle
        =
        \vert
        \{
        n_l^\lambda
        \}         \rangle
$,
$l$ labeling the different momenta. We therefore simply
have
$        \langle\lambda\vert
        b^{\scriptscriptstyle\dagger}_\eta b_\mu
        \vert\lambda\rangle
        =\delta_{\eta,\mu} n_\mu^\lambda
$
and taking the slow energy dependence of
 ${\tilde{t}}$ into account the
contributions with $\lambda=\xi$ cancel out in the master
equation and we need only consider the primed sum for
$\lambda\neq\xi$.
Using, for $\lambda\neq\xi$,
$
        \langle\lambda\vert
        b^{\scriptscriptstyle\dagger}_\eta b_\mu
        \vert\xi\rangle
        =
        \bigl(
        \prod_{\nu\neq\mu,\eta}
        \delta_{n_\nu^\lambda,n_\nu^\xi}
        \bigr)
        \delta_{(n_\eta^\lambda-1),n_\eta^\xi}
        \delta_{n_\mu^\lambda,(n_\mu^\xi-1)}
        (1-\delta_{\eta,\mu})
        \sqrt{n_\mu^\xi}
        \sqrt{n_\eta^\lambda}
$
we come to, setting 
$
{{\bbox{Q}}_{\mu\eta}}\equiv{\bbox{p}}_{\mu}-{\bbox{p}}_{\eta}
$
\widetext
        \begin{eqnarray*}
        {1 \over \hbar}  
        \sum^{}_{{\lambda,\xi}}
        {\hat {\sf L}}_{\lambda\xi} {\hat {\varrho}}
        {{\hat {\sf L}}{}_{\lambda\xi}^{\scriptscriptstyle  
        \dagger}}
        =
        {2\varepsilon\over\hbar}
        \sum_{p p'}
        \sum_{\eta\mu}{}'
        &&
        \left \langle
        n_\mu(1 \pm n_\eta)
        \right \rangle
        {
        \tilde{t}
        \left(
        {
        \left[
        {\bbox{p}}+{\bbox{Q}}_{\mu\eta}
        \right]^2
        /
       2M
        }
        +i\varepsilon,
        {{Q}_{\mu\eta}}
        \right)
        \over
        {{\bbox{p}}_{\mu}^2\over 2m}
        -
        {({{\bbox{p}}_{\mu} - {\bbox{Q}}_{\mu\eta}})^2\over 2m}
        +
        {
        {{\bbox{p}}}^2
        \over
        2M
        }
        -
        {({{\bbox{p}} + {\bbox{Q}}_{\mu\eta}})^2\over 2M}
        +i\varepsilon
        }
        \\
        &&
        \times
        e^{{i\over\hbar}{{\bbox{Q}}_{\mu\eta}}\cdot{\hat {{\sf x}}}}
        \vert {\bbox{p}} \rangle
        \langle {\bbox{p}} \vert
        {\hat \varrho}
        \vert {\bbox{p}}' \rangle
        \langle
        {
        {\bbox{p}}'
        }
        \vert
        e^{-{i\over\hbar}{{\bbox{Q}}_{\mu\eta}}\cdot{\hat {{\sf x}}}}
        {
        \tilde{t}^*
        \left(
        {
        \left[
        {\bbox{p}}'+{\bbox{Q}}_{\mu\eta}
        \right]^2
        /
        2M
        }
        +i\varepsilon,
        {{Q}_{\mu\eta}}
        \right)
        \over
        {{\bbox{p}}_{\mu}^2\over 2m}
        -
        {({{\bbox{p}}_{\mu} - {\bbox{Q}}_{\mu\eta}})^2\over 2m}
        +
        {
        {{\bbox{p}}'}^2
        \over
        2M
        }
        -
        {({{\bbox{p}}' + {\bbox{Q}}_{\mu\eta}})^2\over 2M}
        -i\varepsilon
        }
        ,
        \end{eqnarray*}
where $M$ denotes the mass of the Brownian particle, whose
position operator is
${\hat {\bbox{\sf x}}}$, while
$m$ is the mass of the gas particles.
The Brownian particle is  immersed in a non degenerate gas, so that
$
        \left \langle
        n_\mu(1 \pm n_\eta)
        \right \rangle
        =
        \left \langle
        n_\mu
        \right \rangle
        (
        1 \pm
        \left \langle
        n_\eta
        \right \rangle)
        \approx
        \left \langle
        n_\mu
        \right \rangle
$.
Considering now the quasi-diagonality of the density matrix,
linked to its slow variability, we substitute in the
T-matrix and the denominators ${\bbox{p}}$, ${\bbox{p}}'$
with the symmetric expression $\frac 12 ({\bbox{p}} +
{\bbox{p}}')$; furthermore we use the variables
$
{\bbox{p}}_{\mu}
$,
$
{\bbox{q}} \equiv {\bbox{Q}}_{\mu\eta}
$, and put into evidence the ratio
$\alpha=m/M$ between the masses, thus coming to
        \[
        {4\pi m\over\hbar}
        \sum_{p p'}
        \sum_{q}{}'
        {
        1
        \over
        q
        }
        \!
        \left |
        \tilde{t}
        \!
        \left(
        {
        \left[
        {
        {\bbox{p}}+
        {\bbox{p}}'
        \over
              2
        }+{\bbox{q}}
        \right]^2
        \!\!
        /  \,
        2M
        }
        +i\varepsilon,
        q
        \right)
        \!
        \right |^2
        \!\!
        \sum_{\mu}
        \langle
        n_\mu
        \rangle
        \delta    \!\!
        \left[
        (1+\alpha)
        q
        +\alpha
        (
        {\bbox{p}}+{\bbox{p}}'
        )
        \! \cdot \!
        {
        {\bbox{q}}
        \over
        q
        }
        -
        2
        {\bbox{p}}_{\mu}
        \! \cdot \!
        {
        {\bbox{q}}
        \over
        q
        }
        \right]
        \!
        e^{{i\over\hbar}{{\bbox{q}}}\cdot{\hat {{\sf x}}}}
        \vert {\bbox{p}} \rangle
        \langle {\bbox{p}} \vert
        {\hat \varrho}
        \vert {\bbox{p}}' \rangle
        \langle {\bbox{p}}' \vert
        e^{-{i\over\hbar}{{\bbox{q}}}\cdot{\hat {{\sf x}}}}
        .
        \]
The anticommutator term can be treated
in an analogous way, so that the final expression for the
dissipative contributions in (\ref{eq}) becomes,
neglecting for simplicity the slow energy dependence of
the T-matrix
        \begin{eqnarray*}
        {4\pi m\over\hbar}
        \sum_{q}{}'
        {
        | \tilde{t} (q) |^2
        \over
        q
        }
        \sum_{\mu}
        \langle
        n_\mu
        \rangle
        \Biggl\{
        &&
        \sum_{p p'}
        \delta       \!\!
        \left[
        (1+\alpha)
        q
        +\alpha
        (
        {\bbox{p}}+{\bbox{p}}'
        )
        \cdot
        {
        {\bbox{q}}
        \over
        q
        }
        -
        2
        {\bbox{p}}_{\mu}
        \cdot
        {
        {\bbox{q}}
        \over
        q
        }
        \right]
        e^{{i\over\hbar}{\bbox{q}}\cdot{\hat {{\sf x}}}}
        \vert {\bbox{p}} \rangle
        \langle {\bbox{p}} \vert
        {\hat \varrho}
        \vert {\bbox{p}}' \rangle
        \langle {\bbox{p}}' \vert
        e^{-{i\over\hbar}{\bbox{q}}\cdot{\hat {{\sf x}}}}
        \\
        &&
        \hphantom{x}
        - {1\over 2}
        \sum_{p}
        \delta        \!\!
        \left[
        (1+\alpha)
        q
        +
        2
        \alpha
        {\bbox{p}}
        \cdot
        {
        {\bbox{q}}
        \over
        q
        }
        -
        2
        {\bbox{p}}_{\mu}
        \cdot
        {
        {\bbox{q}}
        \over
        q
        }
        \right]
        \! \!
        \left \{
        \vert {\bbox{p}} \rangle
        \langle {\bbox{p}} \vert,
        {\hat \varrho}
        \right \}
        \Biggr\}
        .
        \end{eqnarray*}
We can now go over to the continuum limit in
$
{\bbox{p}}_{\mu}
$
and
$
{\bbox{q}}
$, evaluating the integral with respect to
${\bbox{p}}_{\mu}$ in the case of a Boltzmann gas, 
using
$
        \left \langle
        n_\mu
        \right \rangle
        = n \lambda^3_m
         \exp [-\beta ({\bbox{p}}_{\mu}^2 / 2m)]
$,
$\lambda_m$ being the thermal wavelength of the gas
particles, $n$ their density
 and
$\beta=1/(k_{\rm {B}}T)$ giving the temperature dependence.
We 
thus obtain, in the Brownian limit
$\alpha\ll 1$
        \[
        {4\pi^2 m^2 \over\beta\hbar}
        n
        \lambda_m^3
        \int d^3\!
        {\bbox{q}}
        \,  
        {
        | \tilde{t} (q) |^2
        \over
        q
        }
        e^{-
        {
        \beta
        \over
             8m
        }
        {{{q}}^2}
        }
        \Biggl[
        e^{{i\over\hbar}{\bbox{q}}\cdot{\hat {{\sf x}}}}
        e^{-{\beta\over 4M}{\bbox{q}}\cdot{\hat {{\sf p}}}}
        {\hat \varrho}
        e^{-{\beta\over 4M}{\bbox{q}}\cdot{\hat {{\sf p}}}}
        e^{-{i\over\hbar}{\bbox{q}}\cdot{\hat {{\sf x}}}}
        - {1\over 2}
        \left \{
        e^{-{\beta\over 2M}{\bbox{q}}\cdot{\hat {{\sf p}}}}
        ,
        {\hat \varrho}
        \right \}
        \Biggr]
        .
        \]
\narrowtext
To get to the master equation describing quantum
dissipation we want to extract the temperature dependence of
this expression, in the limit of small momentum transfer
$
{q}
$.
We therefore expand
the exponential  operators
up to second order in
${\bbox{q}}$,
which is also equivalent to keeping terms at most
bilinear in the operators
${\hat {{\sf x}}}$ and
${\hat {{\sf p}}}$.
Due to symmetry
properties of the coefficients only terms bilinear in
${\bbox{q}}$
and of the form
$
{\bbox{q}}^2_i
$ ($i$ denoting Cartesian coordinates) need be
retained, so that we have
        \begin{eqnarray*}
        &&
        -
        {2\pi^2 m^2 \over\beta\hbar}
        n
        \lambda_m^3
        \int d^3\!
        {\bbox{q}}
        \,  
        {
        | \tilde{t} (q) |^2
        \over
        q
        }
        e^{-
        {
        \beta
        \over
             8m
        }
        {{{q}}^2}
        }
        \sum_{i=1}^3
        {\bbox{q}}^2_i
        \times
        \\
        &&
        \biggl\{
        {
        1
        \over
        \hbar^2
        }
        \left[
        {\hat {{\sf x}}}_i ,
        \left[
        {\hat {{\sf x}}}_i , {\hat \varrho}
        \right]
        \right]
        +
        {
        \beta^2
        \over
         16 M^2
        }
        \left[
        {\hat {\mbox{\sf p}}}_i ,
        \left[
        {\hat {\mbox{\sf p}}}_i , {\hat \varrho}
        \right]
        \right]
        +
        {i\over\hbar}
        {
        \beta
        \over
        2M
        }
        \left[
        {\hat {{\sf x}}}_i ,
        \left \{
        {\hat {\mbox{\sf p}}}_i ,{\hat \varrho}
        \right \}
        \right]
        \biggr\}.
        \end{eqnarray*}
Let us note that in the
derivation important cancellations and compensations arise
between the terms coming from the anticommutator and
incoherent part of (\ref{eq}), necessary in order to obtain
the final structure, thus confirming the fact that {\it in
quantum theory we cannot have separate friction and
diffusion terms}~\cite{LindbladQBM}.
Decisive for the determination of the final structure of the
equation is also the Brownian limit $\alpha=m/M \ll 1$.
Supposing without loss of generality the medium to be
isotropic, so that
$
{\bbox{q}}^2_i =
\frac 13 q^2
$, we can define the
following coefficient 
        \begin{equation}
        \label{gamma}  
        D_{pp}
        =
        \frac 23
        {\pi^2 m^2 \over\beta\hbar}
        n
        \lambda_m^3
        \int d^3\!
        {\bbox{q}}
        \,  
        {
        | \tilde{t} (q) |^2
        }
        q
        e^{-
        {
        \beta
        \over
             8m
        }
        {{{q}}^2}
        }
        \end{equation}        
depending on the collision
cross-section through the T-matrix
and obtain the compact expression
        \[
        \sum_{i=1}^3
        \left \{
        {
        D_{pp}
        \over
         \hbar^2
        }
        \left[  
        {\hat {{\sf x}}}_i,
        \left[  
        {\hat {{\sf x}}}_i,{\hat \varrho}
        \right]  
        \right]  
        +
        {
        D_{qq}
        \over
         \hbar^2
        }
        \left[  
        {\hat {\mbox{\sf p}}}_i,
        \left[  
        {\hat {\mbox{\sf p}}}_i,{\hat \varrho}
        \right]  
        \right]  
        +{i\over\hbar}
        \gamma
        \left[  
        {\hat {{\sf x}}}_i ,
        \left \{  
        {\hat {\mbox{\sf p}}}_i,{\hat \varrho}
        \right \}  
        \right]
        \right \}
        \]
where
$D_{qq} = (
        {  
        \beta\hbar
        /
             4M
        })^2 D_{pp}
$
and
$
\gamma =
        (
        {
        \beta
        /
             2M
        }
        ) D_{pp}
$.
Exploiting (\ref{ansatz}) for the T-matrix we simply obtain
for the potential term in the continuum limit
$
        {\hat {\mbox{\sf V}}}
        =
        -  
        n
        {  
        2\pi \hbar^2  
        \over  
        m  
        }  
        \int d^3\! {\bbox{p}}
        \vert {\bbox{p}} \rangle
        \langle {\bbox{p}} \vert
        \>
        {\mbox{Re}} f(E_p,\theta=0)
$,
$n$ being the density of the gas particles, so that it
essentially depends on the forward scattering amplitude 
$f(E_p,\theta=0)$~\cite{Cohen} as
expected, and vanishes if the latter does not depend on energy. 
The complete master equation then becomes
        \begin{eqnarray}
        \label{qbm}
        {  
        d {\hat \varrho}  
        \over  
                dt  
        }  
        =
        &-&
        {i\over\hbar}
        [
        {{\hat {\mbox{\sf H}}}_0}
        + {{\hat {\mbox{\sf V}}}}
        ,{\hat \varrho}
        ]
        -
        {
        D_{pp}  
        \over
         \hbar^2
        }
        \sum_{i=1}^3
        \left[  
        {\hat {{\sf x}}}_i,
        \left[  
        {\hat {{\sf x}}}_i,{\hat \varrho}
        \right]  
        \right]  
        \nonumber \\
        &-&
        {
        D_{qq}
        \over
         \hbar^2
        }
        \sum_{i=1}^3
        \left[  
        {\hat {\mbox{\sf p}}}_i,
        \left[  
        {\hat {\mbox{\sf p}}}_i,{\hat \varrho}
        \right]  
        \right]  
        -
        {i\over\hbar}
        \gamma
        \sum_{i=1}^3
        \left[  
        {\hat {{\sf x}}}_i ,
        \left \{  
        {\hat {\mbox{\sf p}}}_i,{\hat \varrho}
        \right \}  
        \right]      .
        \end{eqnarray}
This is the main result of this Letter, a CP
time evolution for a quantum Brownian particle
derived at a fundamental level, using a different, new
approach with respect to the usual independent oscillator
model. The equation obtained is translationally invariant and
has the correct Lindblad
form~\cite{LindbladQBM,AlbertoQBM}.
In particular the requirement of CP amounts to check
that
$
D_{pp}D_{qq} \geq
{\hbar^2 \gamma^2 / 4}
$, 
which in our case is verified with the equal sign, thus uniquely
determining the different coefficients in a structure like
(\ref{qbm}) apart from an overall multiplying factor. Let us note that
the requirement of a stationary thermal equilibrium solution only
determines the ratio between $D_{pp}$ and $\gamma$, and also
CP simply indicates that the coefficient $D_{qq}$ should be
different from zero and within some range, without actually fixing
it. This explains the wide variety of different contributions that
have been added to the Caldeira equation to make it preserve
positive definiteness. The fact that 
$
D_{pp}D_{qq} = 
{\hbar^2 \gamma^2 / 4}
$
has as a consequence the following interesting distinctive feature: in
order to write (\ref{qbm}) in a manifest Lindblad form, only one
generator for each Cartesian direction has to be introduced, instead
of two.
In fact using the thermal wavelength
$
\lambda_{M}=\sqrt{\hbar^2 / MkT}
$
associated to the Brownian
particle and defining the operators
$
{\hat {\mbox{\sf a}}}_i=
{
\sqrt{2}
\over
 \lambda_M
}
\left(
{\hat {\mbox{\sf x}}}_i
+{i\over\hbar}
{
\lambda_M^2
\over
                    4
}
{\hat {\mbox{\sf p}}}_i
\right)
$, satisfying
$
\left[
{\hat {\mbox{\sf a}}}_i  ,
{\hat {\mbox{\sf a}}}_j^{\scriptscriptstyle\dagger}
\right]
=\delta_{ij}$, we can rewrite (\ref{qbm}) in the form
        \begin{eqnarray}
        \label{generator}
        {  
        d {\hat \varrho}  
        \over  
        dt
        }  
        =  
        &-&
        {i\over\hbar}
        [
        {{\hat {\mbox{\sf H}}}_0}
        + {{\hat {\mbox{\sf V}}}}
        ,{\hat \varrho}
        ]
        -
        {
        D_{pp}  
        \over
         \hbar^2
        }
        {
        \lambda_M^2
        \over
                 4
        }
        \sum_{i=1}^3
        \frac i\hbar
        \left[  
        \left \{  
         {\hat {{\sf x}}}_i , 
        {\hat {\mbox{\sf p}}}_i
        \right \}            
        ,{\hat \varrho}
        \right]
        \nonumber
        \\
        &+&
        {
        D_{pp}
        \over
         \hbar^2
        }
        \lambda_M^2
        \sum_{i=1}^3
        \left[  
        {
        {\hat {\mbox{\sf a}}}_i
        {\hat \varrho}
        {\hat {\mbox{\sf a}}}_i^{\scriptscriptstyle\dagger}
        - {\scriptstyle {1\over 2}}
        \{
        {\hat {\mbox{\sf a}}}_i^{\scriptscriptstyle\dagger}
        {\hat {\mbox{\sf a}}}_i
        ,{\hat \varrho}
        \}
        }
        \right]      .
        \end{eqnarray} 
This makes an important qualitative difference with a more
phenomenological model derived by Di\'osi~\cite{Diosi}, also linked to
the fact that he obtains an equation with  the asymmetric expression
${\bbox{p}}_\mu$ instead of the momentum transfer ${\bbox{q}}$.
This connection between number of generators and relationships
among the coefficients in a master-equation of the form (\ref{qbm}) 
has not been stressed in the literature, even though it
provides an important qualitative feature, helpful in providing
clearcut distinctions. In this spirit our work also sheds some light
on the recent phenomenological work of Gao~\cite{Gao} and the
subsequent following debate~\cite{Wiseman,Ford}.
Gao works from the very beginning with a single generator
$
V=\mu    {\hat {{\sf x}}} + i \nu       {\hat {\mbox{\sf p}}}
$
and this automatically leads him to obtain a generalized Caldeira
equation with
$
D_{qq} ={\gamma / (8M k_{\rm {B}}T)}
$, so that 
$
D_{pp}D_{qq} = 
{\hbar^2 \gamma^2 / 4}
$
is verified. This explains the difference from the diffusion
coefficient
$
D_{qq} ={\gamma / (6M k_{\rm {B}}T)}
$
found in~\cite{Diosi1-Diosi2}.
Despite the fact that the coefficients in the master-equation with
which Gao starts are actually completely fixed by the requirement of
thermal equilibrium and his choice of a single generator, our work
provides some fundamental evidence in favor of this structure, giving
through  (\ref{gamma}) the quantitative estimate
$
        \gamma =
        (        \beta
        /
             2M
        )
       D_{pp}
$ for the relaxation coefficient.
The heavily criticized~\cite{Wiseman} Hamiltonian term
$
- {\gamma \over 2}         
\left \{  
         {\hat {{\sf x}}}, 
        {\hat {\mbox{\sf p}}}
        \right \}    
$
which Gao obtains, however, does not appear in  (\ref{qbm}), so that
no fictitious counterterm is necessary:
its appearance in rewriting  (\ref{qbm})
in the form  (\ref{generator}) clarifies why the initial choice of 
Gao led to this trouble.
\par
We have thus obtained a new fundamental derivation of
CP dissipative evolution, driven by
collisions with the environment, with temperature dependent
friction and diffusion coefficients expressed in terms of
physical quantities such as the collision cross-section. 
The associated master-equation has the peculiarity of being
expressible in Lindblad form with only a single generator for each
Cartesian direction, thus giving some evidence in favor of a recent
phenomenological model~\cite{Gao}, though being deprived of its
unphysical features~\cite{Wiseman,Ford}.
The
underlying calculations, even though recovering a single
particle description by tracing over matter, are rooted in a
second quantization formalism conceived for the description
of a subset of reduced degrees of freedom slowly varying on
a given time scale~\cite{art1,berlin-goslar}.
A major extension of this model could consist in considering
the effect of correlated initial conditions on the dynamics
and on the property of CP, as we intend to
do in the future.
\par
I am very indebted to Prof. L. Lanz who followed the
whole work and I'd
like to thank Prof. A. Barchielli and Prof. O. Melsheimer
for useful discussions.
This work was supported by the Alexander von
Humboldt-Stiftung.
%%%%%%%%%%%%%%%%%%%%%%%%%%%%%%%%%%%%%%%%%%%%%%%%%%%%%%%%%%%%%
  

\begin{references}  
\bibitem{Dekker-Grabert}
H.~Dekker,
{Phys.~Rep.}
{\bf 80},
1  
(1981);
{H.~Grabert, P.~Schramm, and G.-L.~Ingold},
{Phys.~Rep.}
{\bf 168},
115
(1988).
\par
\bibitem{Alicki}
{R.~Alicki and K.~Lendi},
{\it Lect.~Notes in Physics}, Vol.~286
(Springer, Berlin, 1987).
\par
\bibitem{Kiefer}
{D. Giulini~{\it et al.}},
{\it Decoherence and the Appearance of a Classical World in
Quantum Theory}
(Springer, Berlin, 1996).
\par
\bibitem{Haroche}
{M.~Brune~{\it et al.}},
{Phys.~Rev.~Lett.}
{\bf 77},
4887
(1996).
\par
\bibitem{Lindblad}
{G.~Lindblad},  
{Commun.~Math.~Phys.}
{\bf 48},  
{119}  
({1976}).  
\par  
\bibitem{Kraus-Hellwig}
K.~Kraus,
{\it Lect.~Notes in Physics}, Vol. 190
({Springer}, {Berlin}, 1983);
K.-E.~Hellwig,
{Int. J. Theor. Phys.}
{\bf 34},
{1467}
(1995).
\par
\bibitem{Piza}
{R.~C.~de~Berr\^{e}do~{\it et al.}},
{Physica Scripta}
{\bf 57},
533
(1998).
\par
\bibitem{Pechukas}
{P.~Pechukas},
{Phys.~Rev.~Lett.}
{\bf 73},
1060
(1994);
{\bf 75},
3021
(1995);
R.~Alicki, {\it ibid.}
{\bf 75},
3020
(1995).
\par
\bibitem{Royer-Suarez}
{A.~Royer},
{Phys.~Rev.~Lett.}
{\bf 77},
3272
(1996);
{A.~Suarez, R.~Silbey, and I.~Oppenheim},
{J.~Chem.~Phys.}
{\bf 97},
5101
(1992).
\par
\bibitem{Hu}
{B.~L.~Hu, J.~P.~Paz, and Y.~Zhang},
{{Phys.~Rev.~D}}
{\bf 45},
2843
(1992).
\par
\bibitem{Caldeira}
A.~O.~Caldeira and
A.~J.~Leggett,
{Physica A}
{\bf 121},
587
(1983).
\par
\bibitem{Ambegaokar-Pechukas2}
V.~Ambegaokar,
{\it Ber.~Bunsenges~Phys.~Chem.}
{\bf 95},
400
(1991).
\par
\bibitem{Diosi1-Diosi2}
{L.~Di\'osi},
{Physica A}
{\bf 199},
517
(1993);
{Europhys.~Lett.}  
{\bf 22},
1
(1993).
\par
\bibitem{Sipe}
{A.~Tameshtit and J.~E.~Sipe},
{Phys.~Rev.~Lett.}
{\bf 77},
2600
(1996).
\par
\bibitem{Munro}
{W.~J.~Munro and C.~W.~Gardiner},
{Phys.~Rev.~A}
{\bf 53},
2633
(1996).
\par
\bibitem{pheno}
{H.~Dekker},
{Phys.~Rev.~A}
{\bf 16},
2126
(1977);
{A.~S\v{a}ndulescu and H.~Scutaru},
{Ann.~Phys. (N.~Y.)}
{\bf 173},
277
(1987);
M.~R.~Gallis,
{Phys.~Rev.~A}
{\bf 48},
1028
(1993);
{A.~Isar},
{Helv.~Phys.~Acta}
{\bf 67},
436
(1994).
\par
\bibitem{art1}  
L.~Lanz and B.~Vacchini,  
{Int. J. Theor. Phys.}
{\bf 36},  
67  
(1997).
\par  
\bibitem{art2}
L.~Lanz and B.~Vacchini,  
{Phys. Rev. A}
{\bf 56},
4826
(1997).
\par
\bibitem{Stegholm}
{S.~Stegholm},
{Physica Scripta}
{\bf 47},
724
(1993).
\par
\bibitem{Diosi}  
{L.~Di\'osi},
{Europhys.~Lett.}  
{\bf 30},  
{63}  
({1995}).  
\par  
\bibitem{berlin-goslar}
L.~Lanz and B.~Vacchini,
{Int. J. Theor. Phys.}
{\bf 37},
545
(1998);
{L.~Lanz and O.~Melsheimer},
in
{\it Lect.~Notes in Physics}, edited by A.~Bohm,
H.-D.~Doebner, and P.~Kielanowski
({Springer}, {Berlin}, 1998){, Vol. 504}, p.345.
\par
\bibitem{LindbladQBM}
G.~Lindblad,
{Rep.~Math.~Phys.}
{\bf 10},
393
(1976).
\par
\bibitem{Cohen}
{C.~Cohen-Tannoudji, B.~Diu and F.~Lalo\"e},
{\it Quantum Mechanics}
(John Wiley \& Sons, New York, 1971), Vol.~II.
\par
\bibitem{AlbertoQBM}
A.~Barchielli,
{Nuovo Cimento}
{\bf 74B},
113
(1983).
\par
\bibitem{Gao}
{S.~Gao},
{Phys.~Rev.~Lett.}
{\bf 79},
3101
(1997).
\par
\bibitem{Wiseman}
{H.~W.~Wiseman and W.~J.~Munro},
{Phys.~Rev.~Lett.}
{\bf 80},
5702
(1998); {S.~Gao}, {\it ibid}. {\bf 80},
5703
(1998). 
\par
\bibitem{Ford}
{G.~W.~Ford and R.~F.~O'Connell},
{Phys.~Rev.~Lett.}
{\bf 82},
3376
(1999); {S.~Gao}, {\it ibid}. {\bf 82},
3377
(1999).
\par
\end{references}
\end{document}